# Low temperature dielectric relaxation study of aqueous solutions of diethylsulfoxide

Liana Gabrielyan[1*], Shiraz Markarian[1], Peter Lunkenheimer[2] and Alois Loidl[2]

[1]Department of Physical Chemistry, Yerevan State University, 0025 Yerevan, Armenia

[2]Experimental Physics V, University of Augsburg, 86135 Augsburg, Germany

*E-mail: lgabriel@ysu.am

**Abstract**

In the present work, dielectric spectra of mixtures of diethylsulfoxide (DESO) and water are presented, covering a concentration range of 0.2 - 0.3 molar fraction of DESO. The measurements were performed at frequencies between 1 Hz and 10 MHz and for temperatures between 150 and 300 K. It is shown that DESO/water mixtures have strong glass-forming abilities. The permittivity spectra in these mixtures reveal a single relaxation process. It can be described by the Havriliak-Negami relaxation function and its relaxation times follow the Vogel-Fulcher-Tammann law, thus showing the typical signatures of glassy dynamics. The concentration dependence of the relaxation parameters, like fragility, broadening, and glass temperature, are discussed in detail.

## 1 Introduction

Earlier systematic studies of the solutions of sulfoxides revealed new physicochemical features, which previously have not been reported. For example, it was shown that diethylsulfoxide (DESO), like the widely used dimethylsulfoxide (DMSO), has unique physicochemical properties with possible biomedical applications [1]. Particularly, these systems reveal cryoprotective ability. Recently, the thermal properties of DESO/water systems have been investigated using Differential Scanning Calorimetry (DSC) at rather low heating/cooling rates ($2^{o}$C/min). The glass-forming tendency of these solutions was discussed in terms of existing competing interactions between DESO molecules, on the one hand, and DESO and water molecules, on the other hand [1]. The fact that DESO forms stable glasses over a wider range than DMSO suggests that the glass state may be stabilized by the hydrophobic alkyl surfaces. The cryoprotective ability of DESO on *Escherichia coli* survival was also investigated, and a comparison among DESO and other widely used cryoprotectants, like DMSO and glycerol, was performed [1]. It was shown that DESO compared to DMSO is more effective for preservation of membrane potentials after freezing-thawing of *Escherichia coli* cells. Besides, DESO exhibits more pronounced effects on the anaerobic growth, survival, and ionic exchange of *E. coli* than DMSO [1]. Fourier-transform infrared (FTIR) and Raman, as well as dielectric relaxation studies of pure DMSO, DESO and dipropylsulfoxide in the liquid state and of their solutions in aqueous and non-aqueous solvents at room temperatures suggest a self-associative structure for these compounds [2-4].

In recent years the dynamics of glass formers has attracted considerable interest [5-7]. Dielectric spectroscopy (DS) is an efficient tool to investigate this dynamics [8-10]. Compared to other techniques, DS provides information about time correlations involving the macroscopic polarization [8]. Dielectric relaxation studies of DMSO/water and DESO/water mixtures at room temperature have been the subject of several investigations [3, 11-15]. For the DMSO-water and DESO-water system, permittivity spectra with broad relaxation features at room temperature are obtained that can be described by two Debye-type components [3]. However, single Cole-Cole or Cole-Davidson functions were used for fitting the data as well [3]. In contrast to DMSO [16, 17], dielectric relaxation properties of aqueous solutions of DESO at low temperatures in fact were not investigated in detail so far. In this respect, it should be noted that in [16] the dielectric behavior of the supercooled aqueous solutions of DMSO for the low-frequency range from 20 Hz up to 1 MHz was reported. At temperatures above the glass-transition temperature, it was found that the Havriliak-Negami (HN) relaxation function is appropriate to describe the data [16]. A single relaxation function was also used for the description of dielectric spectra of DMSO/water mixtures in refs. [11, 12, 15, 17].

Our previous investigations using DSC have shown that DESO/water solutions form an amorphous, glassy system at low temperatures, thus avoiding ice crystallization [1]. This is the case for a wide range of concentrations (45 - 75% mass fraction, corresponding to molar fractions 0.12 - 0.34) and even for relatively low cooling rates of 2 K/min. The stability of the amorphous state on heating depends on the DESO content. In order to estimate the stability of this state and to characterize its relaxation dynamics, in the present work we have extended the dielectric measurements of the DESO/water system to low temperatures. Analyzing the temperature-dependent relaxation processes of these systems will add substantial information to the existing knowledge about the glass-forming properties and glassy dynamics of aqueous solutions of sulfoxides in general.

## 2 Experimental Details

DESO was synthesized and purified according to ref. [18]. The solutions were prepared using distilled water. The DESO concentrations $x_{\mathrm{DESO}}$ are specified in molar fractions. The complex dielectric permittivities of the aqueous solutions of DESO were determined for frequencies between 1 Hz and 10 MHz by means of a high-performance frequency-response analyzer (Novocontrol Alpha-A analyzer) [9]. For the measurements the solutions were filled into parallel-plate capacitors with 2.4 mm diameter and 0.55 mm plate distance. For temperature-dependent measurements between 150 K and 300 K, the capacitor containing the sample material was mounted into a $N_2$-gas cryostat (Novocontrol Quatro) [9].

The measurements were done using a moderate cooling/heating rate of 0.4 K/min. Sample temperatures were controlled with a stability of ±0.02 K.

## 3 Results and Discussion

Similar to DMSO/water solutions [16], we found that aqueous solutions of DESO can only be supercooled easily for molar fractions $x_{\mathrm{DESO}}$ ranging from 0.2 to 0.3 (due to the lower cooling rate of 0.4 K/min, this range is somewhat smaller than that obtained from the DSC experiments mentioned above [1]). Therefore the dielectric measurements reported in the present work are limited to this concentration range. Pure DESO shows crystallization at 270 K, the solution with $x_{\mathrm{DESO}} = 0.5$ crystallized at 250 K and $x_{\mathrm{DESO}} = 0.1$ at 226 K. In the present paper, these data are not considered. Raman and FTIR studies of DESO/water mixtures showed that very strong interactions take place between DESO and water, even stronger than those between DESO molecules in the pure liquid and those between DMSO and water in the DMSO/water system [2]. The competing interactions between molecules of DESO, on the one hand, and DESO and water molecules, on the other hand could explain the glass-forming tendency of these solutions.

As a typical example of the temperature variation of the real part $\varepsilon'$ and imaginary part $\varepsilon''$ of the complex dielectric permittivity, fig. 1 shows the results at a frequency of 67 kHz for the sample with $x_{\mathrm{DESO}} = 0.3$. $\varepsilon'(T)$ (fig. 1) shows a continuous increase upon cooling from room temperature and, close to 200 K, a well-pronounced downward step. At the same temperature, the dielectric loss $\varepsilon''(T)$ exhibits a peak. This is the typical signature or relaxational behavior as found in various kinds of dipolar materials [8, 19, 20]. The additional increase of the loss with increasing temperature, occurring above about 225 K, can be ascribed to a conductivity contribution arising from ionic charge transport. It can be assumed to be caused by small amounts of ionic impurities that are almost unavoidable in polar liquids as DESO or water. They lead to non-zero conductivity $\sigma_{dc}$ of the solution and, via the relation $\varepsilon'' \propto \sigma'/\nu$, a corresponding contribution to the dielectric loss arises. An important result of fig. 1 is the absence of any further anomaly, in addition to the relaxation features, that would mark crystallization. Thus, good glass-forming ability of this solution can be stated. Similar results were obtained for the other solutions investigated.

The frequency dependences of the dielectric constant and loss of DESO/water mixture with $x_{\mathrm{DESO}} = 0.3$, measured at various temperatures, are depicted in figs. 2a and b, respectively. The well-pronounced steps in $\varepsilon'(\nu)$, which are accompanied by peaks in $\varepsilon''(\nu)$, again are typical signatures of a relaxational process [8, 9]. They can be ascribed to the structural $\alpha$ relaxation of the investigated solution. As revealed by fig. 2, the position of the relaxation peaks in $\varepsilon''$ and the corresponding steps in $\varepsilon'$ shift toward lower frequency with decreasing temperature. This mirrors the typical increase of the relaxation time due to the glassy freezing of the reorientational motion of the molecules upon cooling [8, 9]. The low-frequency plateau of $\varepsilon'(\nu)$, corresponding to the static dielectric constant $\varepsilon_S$, increases with decreasing temperature as found for most dipolar supercooled liquids [8, 9] and expected within the time-honored Onsager theory [21].The spectra of the other investigated solutions showed similar features.

For an analysis of the dielectric spectra, the HN equation [22],

$$\varepsilon^*(\nu) = \varepsilon_\infty + \frac{\Delta\varepsilon}{\left[1 + (i2\pi\nu\tau)^{1-\alpha}\right]^\beta}, \tag{1}$$

was employed using a nonlinear least-squares fitting procedure. In this empirical formula, $\varepsilon^*(\nu)$ is the complex permittivity, $\Delta\varepsilon$ defines the amplitude of the relaxation process, $\tau$ is the relaxation time of the system, $\varepsilon_\infty$ is the high-frequency limit of the dielectric constant, and $\alpha$ and $\beta$ are the geometric shape parameters defining symmetric and asymmetric broadening of the loss peaks, respectively. By simultaneous fits performed for real and imaginary part of the dielectric permittivity we found that the complex dielectric spectra of DESO/water mixtures can be well described by the HN equation (solid lines in figs. 2a and b). For certain temperatures and concentrations, the $\alpha$ parameter was found to be zero, a case where the HN equation becomes identical with the Cole-Davidson equation [23].

As revealed by fig. 2b, some deviations of fits and experimental data show up at the lowest temperatures and highest frequencies investigated. This finding resembles the so-called "excess wing", which previously was found in a variety of other glass-forming liquids [9, 24-27]. Alternatively, these deviations could be ascribed to a second relaxation process as reported for aqueous solutions of DMSO [16]. In any case, this feature is outside of the scope of the present work and measurements at higher frequencies and/or lower temperatures are needed for a thorough analysis.

Figure 3 shows the imaginary part of the dielectric permittivity as a function of frequency, measured at a temperature of 200 K, for the concentration range $0.2 \leq x_{DESO} \leq 0.3$. Obviously the position of the loss peak shifts toward lower frequency with decreasing DESO concentration implying an increase of the relaxation time. At the same time, with decreasing $x_{DESO}$ a broadening of the spectra and a decrease of the amplitude of the relaxation process is observed. Again the solid lines in fig. 3 correspond to the results of fits using the HN function, eq. (1).

Besides the relaxation time, a dielectric relaxation process is characterized by its strength and broadening. In fig. 4 the temperature dependences of the HN width parameters $\alpha$ and $\beta$ as resulting from the performed fits are shown. The $\beta(T)$ values of DESO/water mixtures vary between about 0.7 and 0.8. At temperatures $T > 220$ K, no significant information on $\beta$ could be obtained as here the high-frequency flank of the relaxation peaks is shifted out of the frequency window (cf. fig. 2). However, by fixing $\beta$ to values deduced from an extrapolation of the data at $T < 220$ K, at least information on $\alpha$ and $\tau$ could be obtained. The $\alpha$ shape parameter, corresponding to a symmetric peak broadening, is close to zero at high temperatures and rapidly increases as the temperature is lowered. While $\beta$ only slightly varies with the composition of the solution, this increase of $\alpha$ is clearly more pronounced for DESO/water mixtures with low content of DESO. Overall, a significant broadening of the relaxation process with decreasing DESO content can be stated as also revealed by fig. 3. Within the heterogeneity scenario, usually invoked to explain the deviations of structural relaxations from monodispersive Debye behavior [28, 29], this is a rather unexpected result as for $x_{\mathrm{DESO}}$ closer to 0.5, naively more disorder and thus stronger heterogeneity may be expected. Interestingly, in aqueous DMSO solutions a similar decrease of $\alpha$ with increasing $x_{\mathrm{DMSO}}$ was found [16]. Moreover, in ref. [11] a minimum in the deviations from a Debye spectral shape was reported for DMSO solutions at a concentration of $x_{\mathrm{DMSO}} \approx 0.33$. This finding was ascribed to the formation of 2:1 water/DMSO aggregates, leading to higher homogeneity of the solutions [11, 16]. One may speculate that a similar mechanism is also active in the present DESO solutions. Indeed the presence of strong interactions between DESO and water molecules leading to molecular aggregates was reported on the basis of Raman and FTIR studies of DESO/water mixtures [2].

The most important parameter obtained from an analysis of relaxation processes is the temperature dependence of the characteristic relaxation time. Instead of the relaxation time $\tau$ defined by eq. (1), molecular interpretations usually refer to the mean relaxation time $\langle \tau_{CD} \rangle$. For the Cole-Davidson distribution ($\alpha = 0$), it is related to $\tau$ by $\langle \tau_{CD} \rangle = \beta\,\tau$ [30]. For the HN function, here we use $\langle \tau \rangle = \tau\beta(1-\alpha)$ as an estimate [31]. Figure 5 provides its temperature dependence in an Arrhenius representation. In ref. [3], $\tau(T)$ data for DESO/water solutions at room temperature were deduced from dielectric measurements in the GHz range. In that work a superposition of up to three Debye peaks was used to parameterize the observed relaxation peak. The average of the reported relaxation times of the two closely overlapping main relaxation peaks is about $10^{-10}$ s. This value is in reasonable agreement with an extrapolation of the present $\tau(T)$ data.

Figure 5 reveals that the relaxation times of the investigated solutions decrease with increasing $x_{\mathrm{DESO}}$. This is in agreement with the shift of the loss peaks documented in fig. 3, implying an acceleration of relaxational dynamics for higher DESO contents. This behavior differs from that in aqueous DMSO solutions. In refs. [11, 15, 16] a continuous slowing down of the relaxation dynamics upon addition of DMSO was reported for concentrations up to about 0.35, followed by a decrease of $\tau$ at higher concentrations. This maximum in $\tau(x_{\mathrm{DMSO}})$ was ascribed to the fact that the binary liquid is "rigidified" compared to the pure liquids, due to hydration effects [11]. In ref. [11], such a maximum was also reported for other aqueous solutions, where it occurred at different molar concentrations. The detected $\tau(x_{\mathrm{DESO}})$ variation of the present DESO/water mixtures in the region $0.2 \leq x_{\mathrm{DESO}} \leq 0.3$ would be consistent with a maximum arising at $x_{\mathrm{DESO}} \leq 0.2$. Moreover, within the above scenario the fact that the relaxation times of the three solutions approach each other upon heating (fig. 5) indicates that the assumed hydration effects become less important at high temperatures. Further investigations at higher concentrations, which, however, only can be supercooled under strong quenching, are necessary to clarify these issues.

The $\langle\tau\rangle(T)$ curves for the investigated DESO/water mixtures show a pronounced curvature in the Arrhenius representation instead of straight-line behavior (fig. 5). In glass-forming liquids, an often used explanation for such deviations from the Arrhenius law, $\tau = \tau_0 \exp(E / k_{\mathrm{B}} T)$ ($E$ denotes an energy barrier), is an increase of the cooperativity of molecular motion at low temperatures [32-34]. The continuous increase of the relaxation times during cooling can be parameterized using the empirical Vogel-Fulcher-Tammann (VFT) equation [35-37], which is the most widespread tool for describing the temperature dependence of the characteristic times of the cooperative dynamics in glass formers [8, 9]. Here we use its modified form as suggested by C.A. Angell [38, 39]:

$$\tau = \tau_0 \exp\left[\frac{D T_{VF}}{T - T_{VF}}\right] \qquad (2)$$

Here, $\tau_0$ is a preexponential factor, $T_{VF}$ denotes the Vogel-Fulcher temperature (Arrhenius behavior is obtained for $T_{VF} = 0$), and $D$ is the strength parameter, used to classify different glass formers within the strong-fragile scheme introduced by Angell and coworkers [38-40]. The lines in fig. 5 are fits of the $\langle\tau\rangle(T)$ curves with the VFT law, eq.(2). The parameters obtained for the investigated DESO/water mixtures are summarized in table 1. The glass-transition temperature $T_g$ for these systems was obtained by extrapolating $\langle\tau\rangle(T)$ to a relaxation time $\langle\tau\rangle(T_g) \approx 100$ s. The obtained glass-transition temperature (table 1) decreases slightly with increasing DESO content. This variation of $T_g$ is consistent with the acceleration of the relaxation dynamics for higher DESO concentrations, discussed above (cf. figs. 3 and 5).

The obtained strength parameters $D$ are in the range from 15 to 17 and a small decrease with increasing DESO concentration is observed (table 1). The strength parameter, related to the so-called fragility, can be used to classify different glass formers according to their deviation from Arrhenius behavior [38, 39]. Small values of $D$ (typically $D < 10$) represent "fragile" behavior showing strong deviations from Arrhenius behavior. For $D$ values >100, the $\tau(T)$ curves are virtually indistinguishable from a straight line in an Arrhenius plot [41]. The obtained values of the strength parameters of the present DESO/water solutions indicate that these mixtures are intermediate between strong and fragile glass formers [41]. From dielectric spectroscopy of a $D_2O$/DMSO mixture with $x_{DMSO} = 0.33$ a strength parameter $D = 15$, of similar magnitude as in the present solutions, was obtained [17]. It should be noted that the strength or fragility of glass formers can be related to the density of minima in the potential energy landscape in configuration space [42, 43]. Thus one may speculate that the increased complexity of the system, induced by the addition of DESO to water, leads to more minima in the energy landscape, thus explaining the found decrease of strength with increasing $x_{DESO}$, documented in table 1.

## 4 Summary and Conclusions

In summary, a detailed dielectric relaxation study of aqueous solutions of DESO in a concentration range of 0.2-0.3 was performed. The dielectric spectra of the investigated solutions show a single relaxation process that reveals the typical signatures of glassy dynamics, i.e., non-Arrhenius slowing down upon cooling and broadening indicating dynamic heterogeneities. The relaxation dynamics of these DESO/water mixtures can be characterized as intermediate within Angell's strong-fragile classification scheme. The concentration dependences of the relaxation parameters partly differ from the behavior in the related DMSO/water system, pointing to differences in molecular associations and hydration. In agreement with previous findings based on DSC measurements, the present work shows that DESO/water mixtures have strong glass forming abilities, which makes them possible candidates for cryoprotectants [1].

## Acknowledgements

A fellowship given to L. Gabrielyan by the German Academic Exchange Service (DAAD) is gratefully acknowledged. This work was partly supported by the Deutsche Forschungsgemeinschaft via Research Unit FOR 1394.

**Table 1.** VFT parameters (eq. (2)) obtained for DESO/water mixtures.

| $x_{DESO}$ | $\tau_0$(s) | $D$ | $T_{VF}$(K) | $T_g$(K) |
|---|---|---|---|---|
| 0.2 | $0.20\times10^{-15}$ | 17.3 | 119.4 | 170 |
| 0.25 | $0.96\times10^{-15}$ | 15.5 | 119.7 | 167 |
| 0.3 | $2.7\times10^{-15}$ | 15.3 | 115.9 | 162 |

**Figure captions**

Fig. 1. Temperature variation of the real part $\varepsilon'$ and imaginary part $\varepsilon''$ of the complex dielectric permittivity at a frequency of 67 kHz for a DESO/water solution with $x_{DESO}$ = 0.3.

Fig. 2. Dielectric constant (a) and dielectric loss (b) spectra of DESO/water mixture ($x_{DESO}$ = 0.3) measured at various temperatures. The solid lines are fits using the HN function, eq. (1).

Fig. 3. Comparison of dielectric-loss spectra of DESO/water mixtures at three concentrations ($x_{DESO}$ = 0.2, 0.25 and 0.3), measured at a temperature of 200 K. The solid lines are fits using the HN function, eq. (1).

Fig. 4. Temperature dependence of the width parameters $\alpha$ and $\beta$ for DESO/water mixtures obtained from the simultaneous fits of the real and imaginary parts of the complex dielectric permittivity with the HN function, eq. (1).

Fig. 5. Temperature dependence of the average relaxation times of DESO/water mixtures obtained from fits of the permittivity spectra.

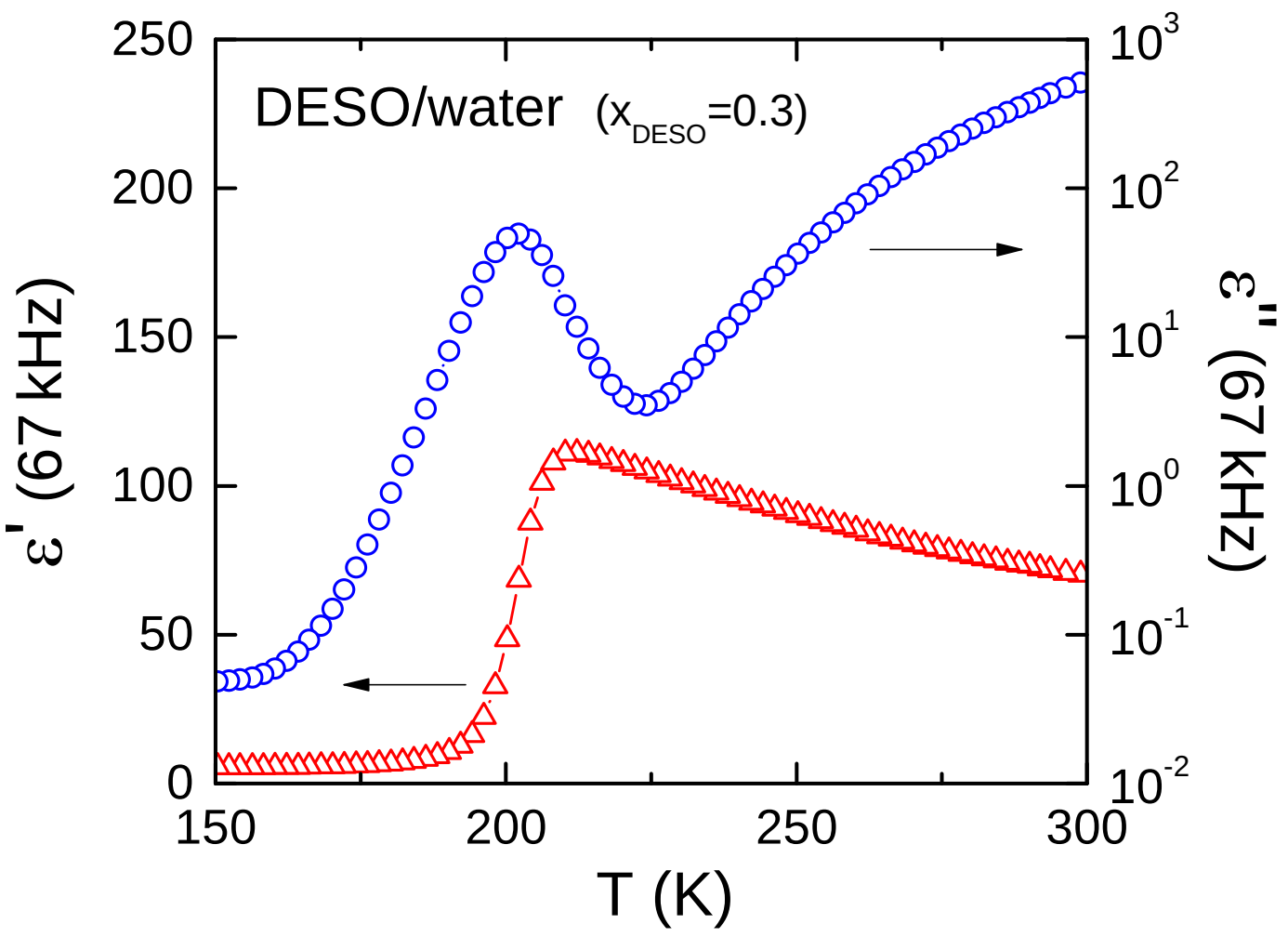

DESO/water (x_DESO=0.3)
ε' (67 kHz)
ε'' (67 kHz)
T (K)
250
200
150
100
50
0
150
200
250
300
10^3
10^2
10^1
10^0
10^-1
10^-2

Fig. 1.

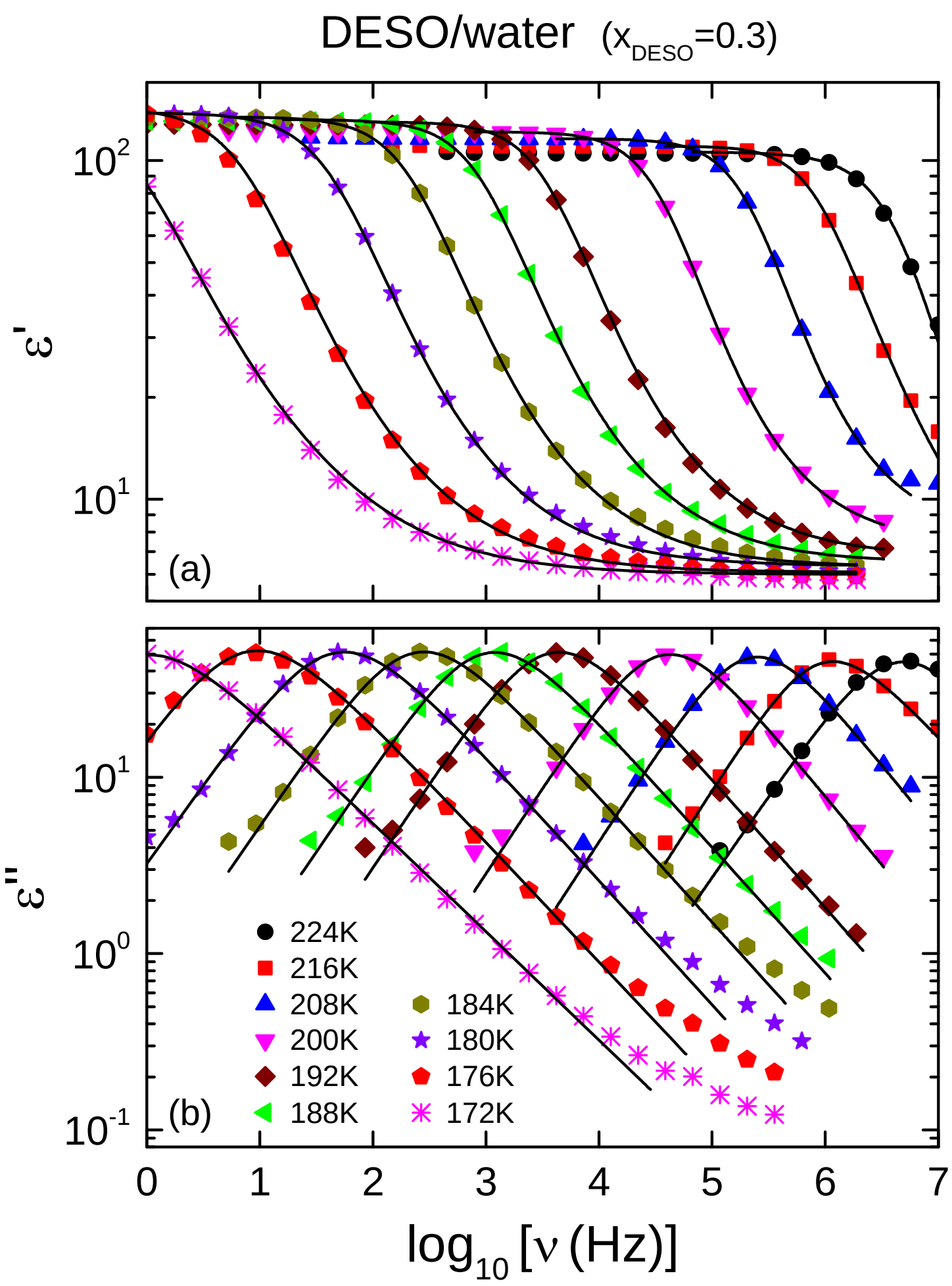
DESO/water ($x_{DESO}$=0.3)
ε'
ε''
(a)
(b)
$10^2$
$10^1$
$10^0$
$10^{-1}$
224K
216K
208K
200K
192K
188K
184K
180K
176K
172K
0
1
2
3
4
5
6
7
$\log_{10}$ [ν (Hz)]

Fig. 2.

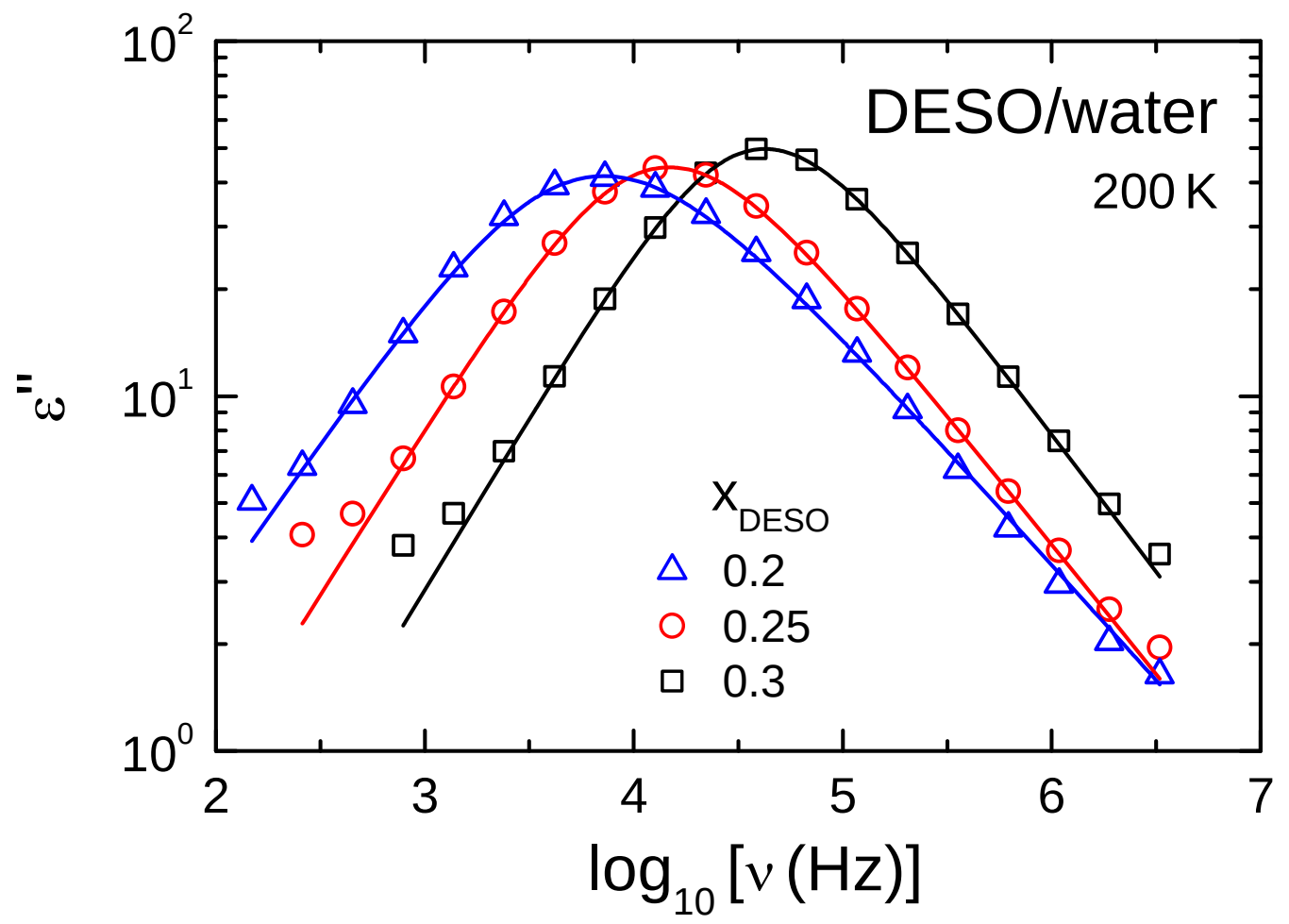

Fig. 3.

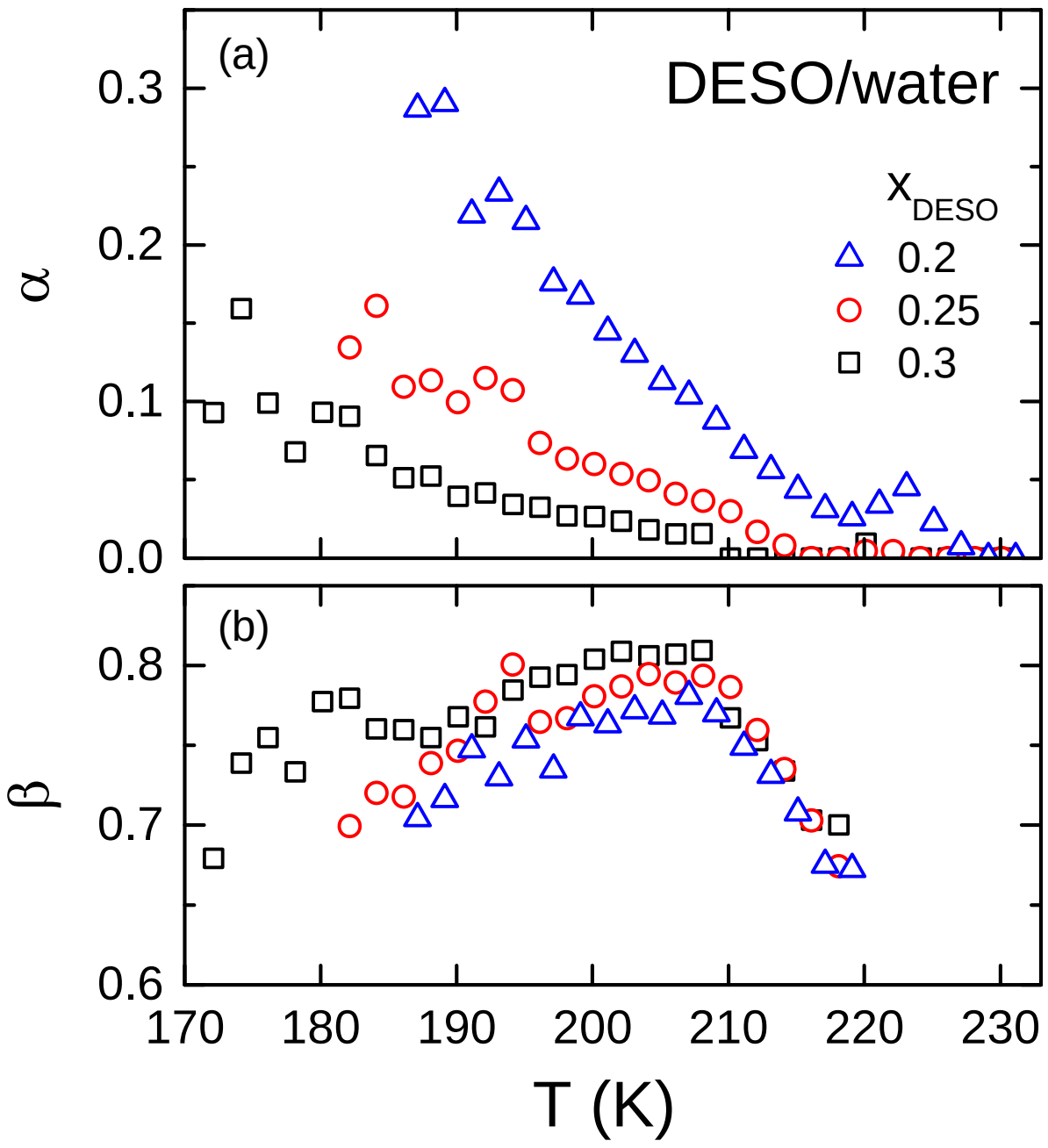

(a)
DESO/water
$x_{DESO}$
0.2
0.25
0.3
α
0.0
0.1
0.2
0.3
(b)
β
0.6
0.7
0.8
170
180
190
200
210
220
230
T (K)

Fig. 4.

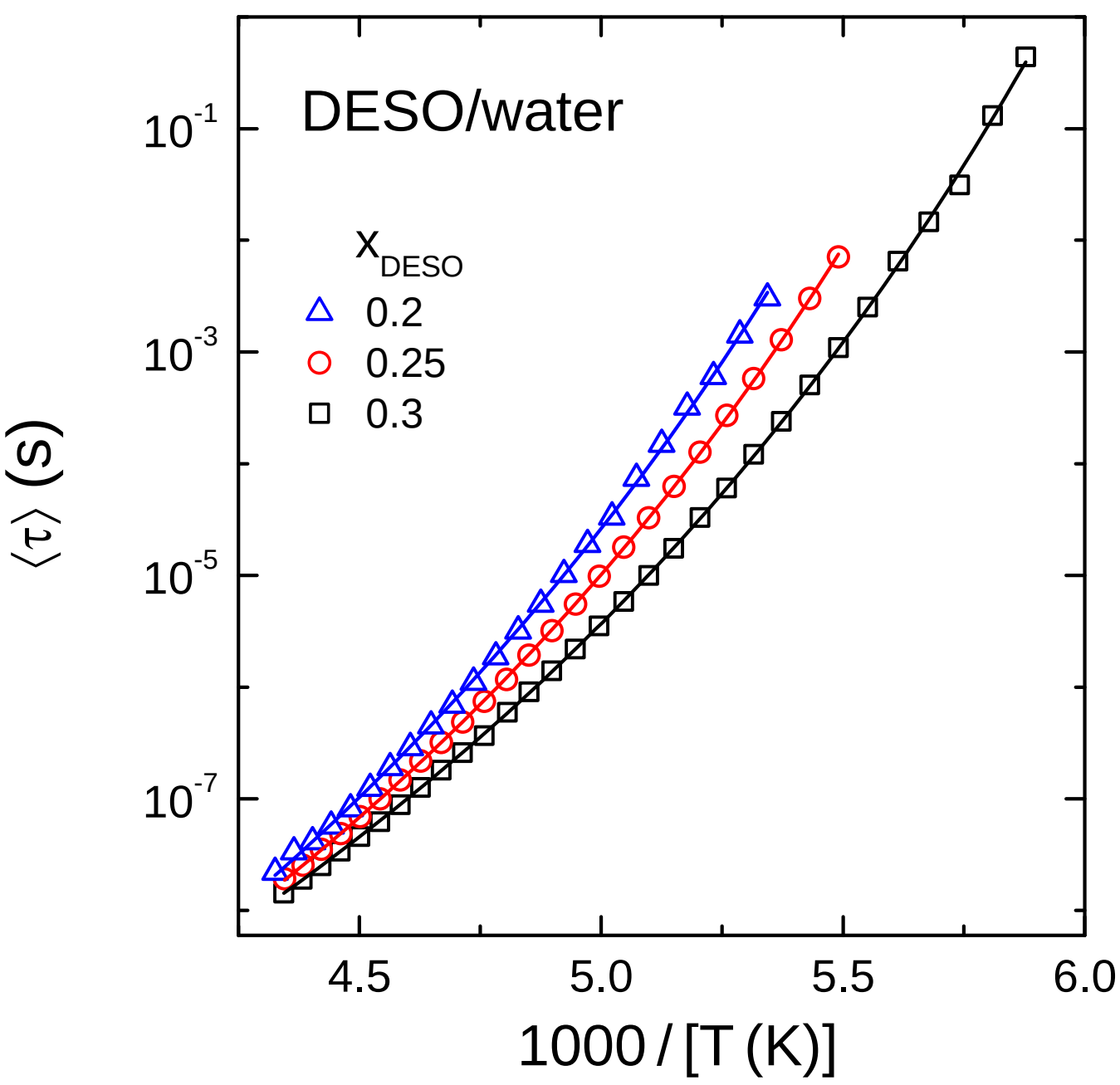
DESO/water
$x_{DESO}$
0.2
0.25
0.3
$10^{-1}$
$10^{-3}$
$10^{-5}$
$10^{-7}$
$\langle\tau\rangle$ (s)
4.5
5.0
5.5
6.0
1000 / [T (K)]

Fig. 5.